\documentclass{vgtc}                          

\graphicspath{{figures/}{pictures/}{images/}{./}} 

\usepackage{times}                     

\usepackage{tabu}                      
\usepackage{booktabs}                  
\usepackage{lipsum}                    
\usepackage{mwe}                       

\usepackage{mathptmx}                  

\onlineid{0}

\vgtccategory{Research}

\vgtcinsertpkg

\title{A Comparative Assessment of Technology Acceptance and Learning Outcomes in Computer-based versus VR-based Pedagogical Agents}

\author{Aimilios Hadjiliasi\thanks{AHadjiliasi1@uclan.ac.uk}\\ %
        \scriptsize UCLan Cyprus %
\and Louis Nisiotis\thanks{LNisiotis@uclan.ac.uk}\\ %
     \scriptsize UCLan Cyprus%
\and Irene Polycarpou\thanks{IPolycarpou@uclan.ac.uk}\\ %
     \parbox{1.4in}{\scriptsize \centering UCLan Cyprus}}

\abstract{
      As educational technology evolves, the potential of Pedagogical Agents (PAs) in supporting education is extensively explored. Typically, research on PAs has primarily focused on computer-based learning environments, but their use in VR-based environments and integration into education is still in its infancy. To address this gap, this paper presents a mixed method comparative study that has been conducted to evaluate and examine how these computer-based PAs and VR-based PAs compare, towards their learning efficacy and technology acceptance. 92 Computing and Engineering undergraduate students were recruited and participated in an educational experience focusing on computing machinery education. The findings of this study revealed that both approaches can effectively facilitate learning acquisition, and both technologies have been positively perceived by participants toward acceptance, without any significant differences. The findings of this study shed light on the potential of utilizing intelligent PAs to support education, contributing towards the advancement of our understanding of how to integrate such technologies to develop learning interventions, and establishing the foundation for future investigations that aim to successfully integrate and use PAs in education.
} 


\begin{document}
\firstsection{Introduction}
\maketitle
Over the past decade, education has witnessed rapid evolution, particularly in the field of immersive learning, due to technology’s advancements. Within these lines, Virtual Reality (VR) and Pedagogical Agents (PAs) have emerged as promising tools that can support teaching and learning. However, while an extensive body of research has been conducted in exploring these two technologies individually, research on PAs in VR is still in its infancy and the potential of such technology in supporting education, is yet to be explored \cite{dai2022systematic, petersen2021pedagogical, siegle2020immersive}.  Building on these lines, we believe that there is no more relevant time to investigate and shed light on the extent to which the use of PAs in VR-based learning environments can support learning. To the best of our knowledge, some research has been conducted to evaluate the acceptance and learning outcomes of using PAs, but only on computer-based systems. With that said, a study has been conducted as a part of a broader research project that seeks to address the subject matter, by exploring several factors that are influencing the successful integration of technology in education. The following paper presents the study, its results and provides a discussion of its findings, contributions and implications.
\section{Background and Related Work}
\subsection{Pedagogical Agents}
PAs are virtual, autonomous, intelligent, embodied and typically life-like characters, inhabiting virtual learning environments, designed to interact, guide and support users during learning experiences, by providing demonstrations, scaffolding and interactive feedback \cite{clarebout2012pedagogical, heidig2011pedagogical}. PAs are typically used to simulate the interaction between a learner and a professional, like an instructor, guide, teacher, mentor, or even a co-learner, and have been utilised in multiple learning domains. In the literature, based on the nature of the environment, the design of PAs varies from a 2D cartoon talking head \cite{veletsianos2012learners} to a 3D humanoid agent capable of displaying emotions, facial expressions and gestures \cite{davis2019sometimes}. In general, research on the effectiveness of PAs still draws blurred results. While some research suggests that the use of PAs can support and enhance students’ learning, other studies revealed that their use can cause students to lose their focus and negatively impact their learning performance. A recent meta-analysis \cite{schroeder2013effective} suggested that the use of PAs has a small but significant effect on learning. Research suggests that the use of PAs is more effective on K-12 students compared to higher education students \cite{tao2022exploring}. Additionally, it was observed that the use of language cues such as speech and written texts, and visual cues such as gaze, emotions and body gestures by PAs, can support students’ cognitive and emotional participation in learning activities and experiences \cite{liew2017exploring}. However, it was also observed that PAs could potentially interfere with students' learning as they can be distracting for students. Although there is not a clear picture considering the most appropriate use of PAs, research indicates that the use of well-designed PAs is more likely to promote deep learning and support teaching and learning \cite{tao2022exploring}.

Typically, research on PAs has been conducted in computer-based desktop environments. It was up until recently that efforts were made to investigate the use and effectiveness of PAs in VR { \cite{dai2022systematic, petersen2021pedagogical, siegle2020immersive}. Initial studies on PAs in VR reported promising results \cite{rickel1998steve}. However, as with technological advancements in agents’ development, PAs do not seem to benefit from the development of VR. Only a few studies evaluated the effects of using PAs in VR, making it hard to conclude how PAs could facilitate learning \cite{dai2022systematic}. Immersive VR has been used for higher education in many different fields \cite{radianti2020systematic} and an adequately designed agent in such an environment could potentially enhance the learning experience by providing the feeling of presence and further strengthening the relationship between the learners and the environment. Thus, the effectiveness of PAs in VR remains an opportunity to be explored, although they might not be beneficial in all cases, due to some challenges/barriers that have to be explored, including technical limitations, associated costs, students’ perceptions, and the need for specialized training for educators, among many others. As a result, the investigation of PAs in VR potentials, challenges and future directions are vital steps in understanding its full educational impact.

\subsection{Technology Acceptance}

Technology acceptance refers to individuals’ willingness and intentions to engage and adopt the use of a new technology. One of the most widely used instruments is the Technology Acceptance Model (TAM) \cite{davis1989perceived}. TAM builds on the idea that is more likely for an individual to accept and use a technology if it is perceived as useful and as easy to use \cite{davis1989perceived}. Due to its applicability, TAM has been extensively used to predict and evaluate the acceptance of different systems in different domains. Following its initial inception, modified versions of TAM have also been proposed to accommodate the advancements in technology and the behaviour of users \cite{putra2018evolution}. Some of these modified versions include TAM2, TAM3 and the Unified Theory of Acceptance and Use of Technology (UTAUT). 

UTAUT builds on TAM and extends on various theories, to explain user experience and behaviours toward using a technology \cite{venkatesh2003user}. UTAUT consists of eight constructs: Performance Expectancy (PE - user's belief that utilizing the technology will enhance their job or task performance), Effort Expectancy (EE - perceived level of ease of users when engaging with a technology), Social Influence (SI - the extent to which users perceive, that others believe they should use the new technology), Facilitating Condition (FC - the degree to which users believe that an organizational and technical infrastructure exists to support the use of the new technology), Self-Efficacy (SE - users' confidence in their ability to use the new technology effectively), and Anxiety (ANX - the degree of apprehension or fear users feel when considering using technology) which are predictors of the Behavioural Intention to Use technology (BITU - individual's intention to use the technology) and Attitudes Toward Using Technology (ATUT – emotional reaction of individuals in engaging with a technology). 

Within the context of PAs, exploring technology acceptance plays an important role in determining their successful integration into education. Several efforts have been made to evaluate their acceptance with different methodology designs mostly focusing on evaluating acceptance of PAs in computer-based systems, but no previous efforts were made to evaluate the acceptance of PAs in VR, and general research on this subject area yields a blurred picture \cite{choi2008comparison, fridin2014acceptance, reich2019acceptance}. This could potentially be due to a lack of sufficient guidelines considering the design and implementation of PAs. Regardless of this blurred picture, investigating PAs’ acceptance into education is important as it can provide us with the necessary insights to ensure their successful and meaningful integration.

\section{Research Approach and Methodology}
To investigate the acceptance and learning efficacy of PAs in VR, a comparative study has been conducted,  comparing its use with the use of PAs in computer-based systems (as a point of reference since we have a certain understanding of their potential). To guide this investigation the following research questions have been formulated:

\textbf{RQ1:}\textit{ How do PAs in VR impact knowledge acquisition in immersive learning activities compared to computer-based PAs?} 

\textbf{RQ2:}\textit{ How do participants' perceptions of effort and performance expectancy, usage intentions, and attitudes toward using PAs in VR compare to computer-based PAs in immersive learning environments?}

Before exploring these questions, we hypothesised that the use of PAs in VR would generate higher technology acceptance and better learning outcomes compared to the use of PAs in computer-based environments. To collect data and answer the research questions, an experimental approach has been designed using multiple data collection instruments and techniques including pre-test and post-test (a set of 15 multiple-choice questions that examine students' prior knowledge and learning gains), a post-experience questionnaire (comprising the four dimensions of UTAUT questionnaire: EE, PE, BITU, ATUT - the rest of the dimensions were not related to the technology’s features. The items were measured in 7-point Likert Scale) and structured focus group discussions (including questions that are related to the acceptance of technology and students perceptions toward its learning efficacy). To conduct this comparison, two identical learning environments have been designed, in VR and 3D desktop, populated with learning materials and PAs.

\subsection{Prototype Environment and Pedagogical Agents}
The learning environment designed to conduct this comparative study represented a virtual museum dedicated to promoting learning on ENIAC, the world’s first electronic computer. The environment featured a 3D replica of ENIAC and its components, and participants were able to closely examine and interact with them to learn about their functions and importance. To support information delivery, facilitate user-environment interactions, and contribute to the overall intelligence of the system, four PAs were implemented, named Mathew, John, George, and Guy. For the design and implementation of the PAs, a thorough examination of the literature has been conducted to identify their capabilities. Building on the findings of this review and in combination with the suggestions of Tao et al. \cite{tao2022exploring} about the PA design characteristics, the two environments and the 4 PAs were developed. Each PA has its specialisation and subsequently different teaching goals. By leveraging symbolic and game AI techniques, these PAs reason their actions and make informed decisions on how to achieve their goals. Considering their roles within the environment, Mathew and John were instructors and specialists in fundamental concepts of ENIAC, its components, and its subsequent functionalities. Their goal was to interact with the user and share informative presentations including an in-depth historical overview of ENIAC’s origin and explanations of its main components’ functionalities, uses, advancements and limitations. George had the role of a teaching assistant. His goal was to provide learners with additional assistance by providing answers to different questions about ENIAC that might arise, after the presentations of Mathew and John. Guy had the role of evaluator and was responsible for assessing users' learning through four different Q\&As, that were embedded within the environment and provided learners with adaptive feedback. To communicate information and provide guidance, feedback and instructions, all PAs were using pre-recorded human voice. Additionally, all agents were using gaze, gestures, body movements and facial expressions as a layer of social cues, which were coordinated with their natural language communication. All agents featured spatial awareness, allowing them to navigate within the environment, decide the shortest path to reach destinations and avoid collisions with obstacles and the learners. PAs were implemented with contextual knowledge, which was used to provide them with a certain level of proficiency considering the environment’s materials. They also maintained episodic memory, to recall events (e.g., remember when a presentation is provided and adapt the presentation if the learner wants to re-attend it). PAs were also capable of continually monitoring the state of the environment and learners' progress, to enable them to make informed decisions and reason their actions. To do that, all the PAs were in constant communication. Lastly, PAs had a common human-like appearance, personality, and behaviour, conveying different personality traits, depending on their role in the learning environment.

\subsection{Experimental Procedures}
To conduct the comparative experiments, 92 undergraduate students studying  Computing and Engineering courses at the University of Central Lancashire, Cyprus (UCLan Cyprus) were recruited to participate and were randomly separated into two groups (control and experimental). Participants of the control group (N=46, Male=39, Female=7) experienced the system in 3D desktop mode, and participants of the experimental group (N=46, Male=36, Female=9, Not Specify=1) used the VR system. Before interacting with any of the study materials, participants have given their informed consent. They were then administered a pre-test to capture their initial knowledge of the topics demonstrated in the learning environment. Following this, they engaged in a learning experience using the technology they were assigned for approximately 45 minutes. At the end of the learning experience, they were administered the post-test to capture their knowledge after the experience and a post-experience questionnaire, to capture data considering technology acceptance. To further investigate with qualitative data the use of PAs in VR, experimental group participants were invited to participate in the focus group discussion. Of the 46 participants, 12 expressed interest in participating. A series of focus group discussions were organised on a different day within the same week. Following data collection, the collected data were analysed using descriptive statistics, statistical tests, and thematic analysis. The analysed data were then mixed and compared to answer the research questions.

\section{Results and Discussion:}
To answer the first research question and examine how the two technologies compare in terms of learning outcomes, and consequently explore whether the use of PAs in VR can promote learning gains, several methods including visual inspections, descriptive statistics, statistical tests and thematic analysis were utilised. For the collected data of both groups, at first, the difference value of the pre-post-test was calculated. Then, a normality investigation to examine the distribution of data was conducted in all three variables of both groups, using the Shapiro-Wilk test, due to the relatively small sample of both groups. The test revealed a violation of normality in all three variables. Following, a visual inspection of differences was conducted revealing only positive values, indicating that all participants had learning gains in both groups. To examine whether the learning gains are of statistical significance, the Wilcoxon signed-rank test was conducted (due to not normally distributed data). The test revealed that this difference, for both groups, is of statistical significance, indicating that learning has been effectively achieved in the use of both groups for learning. Following, to examine how these two technologies compare, the sum of all differences was calculated and compared between the two groups. The inspection revealed small differences between the two groups, weighting on the control group side (Control group: $\Sigma$=247; Experimental Group:$\Sigma$=221).  To examine whether this difference is significant, statistical tests have been conducted using the Mann-Whitney U test. The test revealed small but not statistically significant differences, toward the control group side.  These findings indicate that while there are some differences in learning gains, these are likely due to random chance and not due to true effect. To further examine the learning outcomes of using PAs in VR, thematic analysis has been conducted on the data collected from focus groups. The results revealed that its use for learning had a positive impact on participants’ learning experience. Except for one participant who reported experiencing dizziness after some time, participants expressed that the use of technology has positively impacted their learning, fostering enhanced engagement and deep understanding of information delivered in the learning environment when using the unified technology, due to its capability of providing an immersive and interactive method of learning.

\textbf{P3:}\textit{” I only have positive feedback as this technology allows me to easily learn…I am still remembering everything that was presented...”}

\textbf{P5:}\textit{ ”Before...I didn’t know anything about ENIAC and in only 20 minutes I learned everything about it...”}

The findings demonstrate that both technologies can effectively promote students' learning acquisition. Considering the initial hypothesis, the results suggest that there is evidence to reject it as the use of PAs in VR did not outperform computer-based PAs. However, these differences are likely due to random chance. Potentially this is a result of the different affordances of the two technologies. Hence, further research is required to establish these findings. Focusing on VR-based PAs, the findings demonstrate important implications considering its use to support education. This study demonstrates that such technology is effective for promoting knowledge acquisition, allowing educational institutions to use such technology to create learning interventions which are beneficial towards learning to students. These findings align with the findings of Grivokostopoulou et al., \cite{grivokostopoulou2020effectiveness}. The positive feedback for VR-based PAs highlights the potential of their use to further motivate and immerse learners. However, these findings appear to be situational, as this study investigates one learning intervention and not the long-term effect of such technology potential in promoting learning. Hence, future research should explore the long-term impacts of VR-based learning and identify key elements that contribute to its effectiveness. To this end, this investigation suggests that the integration of VR into education could be considered a promising learning tool to create effective learning experiences for students.

To answer the second research question, compare the acceptance of the two technologies and consequently, evaluate the students’ perceptions toward the acceptance of using PAs in VR, descriptive statistics, statistical tests and thematic analysis were utilised. Before any analysis, a test of reliability using Cronbach’s alpha coefficient was conducted on the measured dimensions, revealing high internal consistency for both groups across all the dimensions. Following this, a test of normality was conducted using the Shapiro-Wilk test. The test revealed that all data across all dimensions of both groups violated the assumption of normality, therefore non-parametric statistics were used to evaluate the acceptance of both technologies. Descriptive statistics were measured with median (md) and Interquartile Range (IQR), and the Mann-Whitney U test was employed to determine statistical differences between the groups. The results revealed that participants found both technologies as moderately easy to use (EE)(Control group: md=5.50, IQR=1.13; Experimental group: md=5.75, IQR=1.00), and at the same time they believe that the use of technology will help them to learn easily (PE)(Control group: md=6.00, IQR=1.50; Experimental group: md=6.00, IQR=.50). Furthermore, for both technologies participants expressed very positive intentions and willingness (BITU)(Control group: md=5.00, IRQ=1.00; Experimental group: md=6.00; IQR=1.00), and overall emotional reaction (ATUT)(Control group: md=6.00, IQR=1.63; Experimental group: md=6.00, IQR=1.00) toward the use of these technologies for future learning. Following this analysis, visual inspections were conducted and some differences were observed between the two groups. To examine whether these differences are of statistical significance, the Mann-Whitney U test was utilised. The test revealed that even though there are some minor differences between the two technologies, those are not significant, denoting that their differences are likely due to random chance and not due to true effect. 
To further examine the technology acceptance of using PAs in VR, a thematic analysis has been conducted on the data collected from focus groups. The analysis revealed that the acceptance of this technology appeared to be generally positive among participants across all four investigated dimensions.

\textbf{P1:}\textit{ “I would use it because it is an interactive method of learning...So, if you ask me if I have to use this technology in the future, I will say Yes...”}

\textbf{P2:}\textit{”I could easily use it for future learning...makes my learning more immersive, like a game!...”} 

The findings demonstrate that both technologies were well-received by participants, with VR-based PAs having a slight advantage over computer-based PAs. The results suggest that VR-based PAs outperform computer-based ones to some extent, though not due to a true effect. Similar to learning outcomes, this might be a result of the two technologies' affordances. Thus, considering the overall positive acceptance of the two technologies, and the non-significant differences which have been observed, there is evidence suggesting that VR-based PAs have the potential to be a promising tool for creating positively perceived learning interventions, indicating their potential to transform learning experiences and develop a new and effective learning technology.

\section{Contributions and Implications}
The findings of this research demonstrate several contributions and implications for both research and educational practices. A key contribution is the investigation of PAs in VR as a learning tool, toward its acceptance and learning efficacy. Such investigation advances our understanding towards the potential of such technology in supporting education, specifically within the domain of computer science education. The findings demonstrated that such technology is positively perceived by learners and effectively promotes knowledge acquisition. The results suggest that such technology should be considered as a learning tool that can be used to create meaningful learning interventions. Another significant contribution lies in the investigation of PAs in VR, technology acceptance and specifically within the lines of PE, EE, BITU and ATUT dimensions. The findings demonstrated positive perceptions and high willingness from students, toward the use of this technology for learning. Hence, these findings have insights into the acceptance of such technology and demonstrate that it can promote learning outcomes and is also accepted by students who are willing to use it, have positive attitudes toward its use, and believe that such technology can increase their performance. This implication further supports the idea that this technology should be considered for dynamic and interactive learning interventions to support teaching and learning. This research also sets the premises for future research directions. Future studies could focus on areas that are not captured by this study, such as the application of technology in different populations, the investigation of long-term impacts on learning efficacy and acceptance, as well as research on other factors such as learning engagement and motivation among many others. These are some of the many factors that could be investigated to ensure the successful integration of technology in education. 
\section{Conclusions, Limitations and Future Directions}
This study investigated the use of PAs and VR and explored its learning efficacy and acceptance by Computing and Engineering undergraduate students. The findings suggest promising outcomes considering the technology's efficacy in promoting learning outcomes and demonstrated the positive perceptions of participants toward its acceptance. 
The fusion of PA and VR presents significant implications and potential for transformative educational experiences, offering access to personalized and adaptive learning experiences, with capabilities of enhancing engagement and retention through interactive, and realistic learning interventions. The findings suggest that working with PAs has the potential to enable learners to practice skills and absorb information in safe, controlled environments, fostering access to high-quality, experiential learning across diverse fields. This innovative approach builds the premises for improved technologically enhanced educational practices that accommodate for different learning styles and needs, supporting and revolutionizing traditional learning methodologies for effective and accessible learning experiences.
Despite the contributions of this study, there are some notable limitations which might limit the generalizability and applicability of findings. These limitations are used to shape the future directions of this research. One of the key limitations relates to the sample’s characteristics, limiting the applicability of results to other fields. Future research should involve a broader and more diverse sample, ideally using random sampling. The small sample size of 92 participants divided into two groups of 46 is relatively limited; thus, future studies should aim for larger sample sizes. Another limitation is that the study focused on assessing perceptions and learning outcomes after a single exposure to the technologies, leaving long-term effects unexamined. Future research should explore these technologies’ long-term impacts on knowledge acquisition and acceptance. Additionally, the subjective nature of thematic analysis may introduce bias, as different researchers might interpret qualitative data differently. Future studies should consider these factors to enhance the robustness and generalizability of their findings.

\bibliographystyle{abbrv-doi}

\bibliography{template}
\end{document}